\documentclass[]{jfm}

\usepackage{graphicx}
\usepackage{newtxtext}
\usepackage{newtxmath}
\usepackage{natbib}
\usepackage{hyperref}
\hypersetup{
    colorlinks = true,
    urlcolor   = blue,
    citecolor  = black,
}
\usepackage{booktabs}

\usepackage{bm}

\usepackage[svgnames]{xcolor}
\hypersetup{colorlinks=true, citecolor=NavyBlue, urlcolor=NavyBlue, linkcolor=NavyBlue}

\newcommand{\beq}{\begin{equation}}
\newcommand{\eeq}{\end{equation}}
\newcommand{\beqa}{\begin{eqnarray}}
\newcommand{\eeqa}{\end{eqnarray}}
\newcommand{\tbu}{\tilde{\bf u}}

\def\cD{ \mathcal{D} }

\def\cE{ \mathcal{E} }

\def\cI{ \mathcal{I} }

\def\cF{ \mathcal{F} }

\newcommand{\bu}{{\bf u}} 
 
\newcommand{\bF}{{\bf f}}
\newcommand{\bk}{{\bf k}}
\newcommand{\bq}{{\bf q}}

\newcommand{\bx}{{\bf x}}

\shorttitle{Nonhomogeneous Turbulence}
\shortauthor{A. Alexakis } 

\title{How far does turbulence spread?}

\author{Alexandros Alexakis\aff{1}
  \corresp{\email{alexakis@phys.ens.fr}},
  }

\affiliation{\aff{1} Laboratoire de Physique de l'Ecole Normale Sup{\'e}rieure, ENS, Universit{\'e} PSL, CNRS, Sorbonne Universit{\'e}, Universit{\'e} de Paris, F-75005 Paris, France
}

\begin{document}

\maketitle

\begin{abstract}
How locally injected turbulence, spreads in space is investigated with direct numerical simulations. 
We consider a turbulent flow in a long channel 
generated by a forcing that is localised in space. The forcing is such
that it does not inject any mean momentum in the flow.
We show that at long times a statistically stationary state is reached where the 
turbulent energy density in space fluctuates around a mean profile that peaks at the forcing location and decreases fast away from it.
We measure this profile as a function of the distance from the forcing region  
for different values of the Reynolds number.
It is shown, that as the Reynolds number is increased, it converges to a Reynolds-independent profile implying that turbulence spreads  due to self-advection and not molecular diffusion.
%
In this limit therefore, turbulence plays the simultaneous role of cascading the energy 
to smaller scales and transporting it to larger distances. 
The two effects are shown to be of the same order of magnitude. 
Thus a new turbulent state is reached where turbulent transport 
and turbulent cascade are equally important and control its properties.
%
%
\end{abstract}

\section{Introduction}\label{sec:Introduction}  

A drop of dye in a fluid will spread so that at long times it is uniformly distributed in the entire space.  
This is not necessarily true for a turbulent puff introduced locally in an otherwise still fluid.  
Turbulent energy will also spread either by viscous diffusion or by self-advection but at the same time will dissipate. 
{\it At long times, if constantly injected, will the spreading of turbulence be able overcome the dissipation so that turbulence spreads throughout the domain or dissipation will limit its presence only near its source?  }
The answer to this question is not {\it a priori} obvious and is fundamental for understanding  inhomogeneous turbulent flows.

Inhomogeneous flows have been the subject of various recent studies \citep{valente2011decay,gomes2015energy, portela2020role, araki2022inertial, berti2023mean}
that have all emphasised the effect of inhomogeneity in the cascade process which can make it deviate from the classical homogeneous case. 
In particular it has been shown that inhomogeneity can alter the scale by scale balance of the cascade \citep{apostolidis2022scalings,apostolidis2023turbulent} and change its scaling properties.
Furthermore, inhomogeneity is an indispensable ingredient of many classical canonical flows 
such as
the spreading of a turbulent jet \citep{list1982turbulent,carazzo2006route,ball2012flow,cafiero2019non} and the spreading of turbulence from the boundaries in wall bounded flows 
\citep{
jimenez2012cascades,gomes2015energy,cimarelli2016cascades,mollicone2018turbulence}. 
In these cases  however along with the injection of energy there is also a mean injection of momentum. 
Momentum, unlike energy, is not dissipated by viscosity
and it can only be transferred in space (by viscosity or advection) or out of the domain
through the boundaries by viscous forces. Thus, much like the example of the drop of dye, the 
injected momentum will spread through out the space carrying along energy. 
The same holds if the injected energy has a mean angular-momentum that is also conserved by viscous forces. 
Therefore, in the case that there is mean momentum injection the answer to the question posed in the first paragraph is that momentum and energy will occupy the entire domain. 
The present work investigates the spreading of turbulence in the absence of mean momentum and angular momentum injection which is fundamentally different from the cases mentioned before. 

To do that we consider turbulence generated in a long triple-periodic channel.
The flow is forced homogeneously in the two short directions of the channel 
and locally in the long direction. The forcing is such that no mean momentum is injected.
We study the behavior of the flow inside and outside the forcing region at long times,
measuring the energy distribution and energy fluxes in real and spectral space.
 In the next section \ref{sec:setup} we present the mathematical
set up of the system under study and define all relative quantities under investigation.
In 
section \ref{sec:dns} we present the results from numerical simulations.
Conclusions are drawn in the final section \ref{sec:conclusions} where directions for 
future research are also discussed.

\section{ Formulation}\label{sec:setup} \label{sec:setup}
\subsection{Mathematical setup} 

\begin{figure}                                                                        
  \centerline{\includegraphics[width=0.9\textwidth]{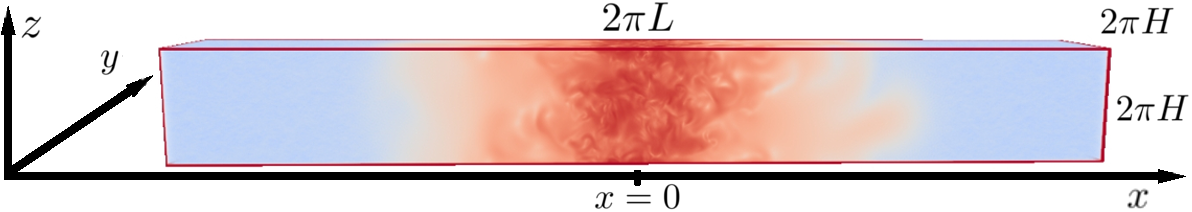}}
  \caption{The computational domain considered. The length $L$ wwas chosen to be eight times the height $L=8H$ and $x=0$ is taken to be at 
  the middle of the channel. The colors indicate visualisations of the enstrophy
  $(\nabla\times \bu)^2$ with red indicating high values while blue are small values.}
\label{fig:ka}
\end{figure}

A triple periodic domain of size $2\pi L \times 2\pi H \times 2 \pi H$ is considered
as shown in figure \ref{fig:ka} with $L\gg H$ being along the $x$-direction and $x=0$ 
is taken to be the mid-plane of the channel. 
The flow inside the domain satisfies the Navier-Stokes equation 
\begin{equation}
    \partial_t \bu + \bu \cdot \nabla \bu = -\nabla P +\nu \nabla^2 \bu + \bF
\end{equation}
where $\bu$ is the divergence free velocity field $(\nabla \cdot \bu=0)$, $P$ is the pressure, $\nu$ is the viscosity and $\bF$ is the forcing.
The functional form of the forcing is given by 
\beq
\bF(t,\bx) = \left[ \begin{array}{c}
     0  \\
     \partial_z [\psi(t,\bx/\ell)-\psi(t,-\bx/\ell)] \\
     \partial_y [\psi(t,-\bx/\ell)-\psi(t,\bx/\ell)]
\end{array} \right] \exp\left[ \frac{L^2}{\ell^2}\left(\cos\left(\frac{x}{L}\right) -1\right) \right] 
\label{eq:forcing}
\eeq 
where $\psi(t,\bx/\ell)$ is a random function including 
only Fourier modes with wave-vectors $\bf k$ satisfying $0<|\bk \ell|\le 2$ and $k_x\ne0$. 
The phases of these modes are delta correlated in time so that the mean energy injection rate is fixed to $\cI_0$. 
The forcing is anti-symmetric with respect to reflections in the $x=0$ plane.
As a result there is zero momentum injection for every realisation.
Furthermore the forcing satisfies $\nabla \cdot \bF=0$.
For $|x|\ll L$ the exponential factor to the right of \ref{eq:forcing} scales like $\exp(-x^2/\ell^2)$ 
so that the forcing is limited only around the range $|x|\sim \ell$ and zero outside. 
In the numerical simulations that follow
we have picked $\ell=H$ and $L=8H$ that was proven (a posteriori) to be long enough so that the effect of the periodicity along the x direction does not play a role.

\subsection{Energy balance relations and fluxes in space}

The primary quantity of interest in this work is the time and volume averaged 
energy density of the system that is given by 
\beq
\cE_0 = \frac{1}{2} \left\langle \left\langle |\bu|^2\right\rangle_{_{V}} \right\rangle_{_{T}} 
\label{eq:energy}
\eeq 
where the angular brackets $\langle\cdot \rangle_{_{T}}$ stand for time average and $\langle\cdot \rangle_{_{V}}$ for volume average defined as
\beq
\langle f \rangle_{_{T}} = \lim_{T\to\infty} \frac{1}{T} \int_0^T f dt\quad \mathrm{and} \quad
\langle f \rangle_{_{V}} =  \frac{1}{V}  \int_V f \,dx\,dy\,dz,
\eeq 
with $V=(2\pi)^3H^2L$ being the system volume.  
The averaged rate $\cI_0$ that energy  is injected is balanced by the averaged rate $\cD_0$ that energy is dissipated, 
leading to 
\beq 
\cI_0 \equiv
\left\langle \left\langle \bu \cdot \bF \right\rangle_{_{T}} \right\rangle_{_{V}} = 2\nu \left\langle  \langle {\bf S}|^2  \rangle_{_{T}} \right\rangle_{_{V}} \equiv \cD_0 
\label{eq:enbal} 
\eeq 
where $\bf S$ stands for the strain tensor 
\beq 
S_{i,j} = \frac{1}{2}\left[ \partial_i u_j + \partial_j u_i \right].
\label{eq:stress}
\eeq 

However, neither the time averaged energy, nor its injection nor its dissipation are uniform along the $x$ direction. 
It is thus appropriate to consider  the mean energy density in a subdomain of the periodic box
\beq 
\cE(X) =  \frac{1}{2}\left\langle \left\langle  |\bu|^2 \right\rangle_{_T} \right\rangle_{_X}
\eeq
where $\langle \cdot \rangle_{_X}$stands for the average confined in the sub-box from $x=-X$ to $x=X$:
\beq \langle f \rangle_{_X} = \frac{1}{(2\pi H)^2}  \int_0^{2\pi H} \int_0^{2\pi H} \int_{-X}^X \langle f(x,t)\rangle_{_T} \,dx\,dy\,dz. \eeq 
For $X=\pi L$ the entire box is considered so clearly $\cE(\pi L)=2 \pi L \, \cE_0$.
We also define the local energy density averaged over the planes  $x=\pm X$
\beq 
E(t,X) = 
       \frac{1}{2(2\pi H)^2}\int_0^{2\pi H} \int_0^{2\pi H} |\bu(t,X,y,z) |^2+|\bu(t,-X,y,z) |^2 dydz .
\label{eq:energyd}
\eeq 
The two energy densities are related by
$\left\langle E(X) \right\rangle_{{T}}=\partial_X  \cE(X) $. 
%

A generalisation of eq. \ref{eq:enbal} for $\cE(X)$ can then be obtained by taking 
the inner product of the Navier-Stokes equation with $\bu$ time averaging and integrating over $y,z$ and from $x=-X$ to $x=X$  to obtain: 
\begin{equation}
    \mathcal{I}(X) =  \mathcal{D}(X)   + \mathcal{F}(X) 
\end{equation}
where $\mathcal{I}(X)$ and $\mathcal{D}(X)$ are the energy injection rate the energy dissipation rate 
within the considered volume defined respectively  as:
\begin{equation}
    \mathcal{I}(X) \equiv \left\langle  \left\langle  \bF(t,\bx)\cdot \bu(t,\bx)   \right\rangle_X  \right\rangle_T
    , \quad \mathrm{and} \quad
    \mathcal{D}(X) \equiv 2\nu \left\langle \left\langle   |{\bf S}(t,\bx)|^2   dx \right\rangle_X \right\rangle_T.
\end{equation}
The third term $\cF(x)$ is a flux that expresses the rate energy is transferred outside the considered volume \citep{landau2013fluid}. 
It can be decomposed in three terms 
\begin{equation}
\cF = \cF_U+\cF_P+\cF_\nu 
\label{eq:fluxbal}
\end{equation}
where $\cF_U$ is the energy flux due advection, $\cF_P$ the flux due to pressure and $\cF_\nu$ the flux due to 
viscosity. They are defined explicitly as
\begin{eqnarray}
    \cF_U(x)     &=& \frac{1}{2(2\pi H)^2} \left\langle \int_{x=X} u_x |\bu|^2 dy\,dz - \int_{x=-X} u_x |\bu|^2 dy\,dz  \right\rangle_T, \\
    \cF_P(x)     &=& \frac{1}{(2\pi H)^2}  \left\langle \int_{x=X} u_x P dy \,\,dz- \int_{x=-X}u_x P dy \,\,dz    \right\rangle_T \\
    \cF_\nu(x)   &=& \frac{\nu}{(2\pi H)^2}\left\langle 
                     \int_{{x=-X}}  u_i \partial_i u_x+ u_i \partial_x u_i    dy dz -
                     \int_{{x= X}}  u_i \partial_i u_x+ u_i \partial_x u_i     dy dz
                    \right\rangle_T
\end{eqnarray}
where the integrals are taken  at the two planes $x=\pm X$ and summation over the index $i$ is assumed in the last one.

\subsection{Energy spectra and fluxes in scale space}

The fluxes above describe how energy is transported in physical space. At the same time, 
energy is also transferred in scale space from large to small scales. To quantify the energy distribution and fluxes in scale space we use the Fourier transformed fields
$\tbu_\bk(t)$ defined by
\beq
\tbu_\bk(t) = 
\left\langle \bu(t,\bx) \, e^{-i\bk \cdot \bx} \right\rangle_{_V}
\quad\mathrm{and}\quad 
\bu(t,\bx) = \sum_{\bk} \tbu_\bk(t) e^{i \bk \cdot \bx}, 
\eeq 
where the inverse wavenumber $k^{-1}$ gives a natural definition of a scale.
The energy spectrum, giving the distribution of energy among scales is defined as 
\beq 
\tilde{E}(k) = \frac{1}{2}\sum_{k<|\bq|<k+1} \left\langle |\tbu_\bq|^2 \right\rangle_{_T}
\eeq 
The energy flux gives the rate that energy flows across $k$ is defined as
\beq 
\Pi(k) = -\left\langle \left\langle \bu^<_k \cdot \bu\cdot \nabla \bu \right\rangle_{_V}\right\rangle_{_T}
\eeq
where $\bu^<_k$ stands for the velocity field filtered so that only wavenumbers with norm $|\bk|<k$ are kept \citep{alexakis2018cascades,frisch1995turbulence}.

\subsection{Reynolds numbers}  

The Reynolds number in this system provides a measure of the strength of turbulence 
is typically defined as $ Re ={U\ell}/{\nu}$
where $U$ is the typical velocity of the system.
In this work we are interested in the long box limit $L \gg H$ and some care needs to 
be taken in order to be able to compare with homogeneous turbulence results. 
If we define $U$ based on the mean energy density $\cE_0$ (given in eq. \ref{eq:energy})
then if turbulence remains localized,  $\cE_0$ will approach zero in the limit $L\gg H$.  
Thus defining $U$ as the root mean square (rms) value over the entire domain $U=(2\cE_0)^{1/2}$ will greatly underestimate 
the value of $U$ close to the forcing region. The same holds for the mean dissipation rate density $\cD_0$. 
To compensate for that we will define the typical velocity $U$ and the typical dissipation rate 
$\epsilon$ as
\begin{equation}
    U=\sqrt{\frac{2\cE_0 L}{H}}, \quad 
    \epsilon = 
               \cD_0 \frac{L}{H}
\end{equation}
The factor $L/H$ introduced makes $U$ and $\epsilon$ remain finite in the $L/H\to\infty$ limit for localised turbulence.
These definitions can be interpreted as the rms velocity and dissipation around the forcing region.
With these definitions of $U$ and $\epsilon$ the following three Reynolds numbers typically met in the literature are defined:
\beq 
Re_U        \equiv \frac{UH}{\nu}, \quad 
Re_\epsilon \equiv \frac{\epsilon^{1/3} H^{4/3}}{\nu}, \quad 
Re_\lambda  \equiv \frac{ \sqrt{5} U^2}{(\nu\epsilon)^{1/2}}.
\eeq 
The first one is the classical definition of the Reynolds number based of the (re-scaled) rms velocity.
The second is a Reynolds number based on the energy injection/dissipation and is the one we control in these simulations
(since it is the energy injection rate we impose). Finally the third one is the Taylor-Reynolds number based 
on the Taylor micro scale $\lambda=U\sqrt{5\nu/\epsilon}$. The three definitions are related by 
\beq 
5Re_U^4 = Re_\lambda^2 Re_\epsilon^{3}.
\eeq 
and for large $Re_U$ it is expected that $Re_U\propto Re_\epsilon \propto Re_\lambda^2$. 

\subsection{Numerical Setup}

The Navier-Stokes equations are solved using the pseudo-spectral code 
{\sc ghost} \citep{mininni2011hybrid}, that uses a 2/3 de-aliasing rule
and a second order Runge-Kuta method for the time advancement.
A uniform grid was used such that the grid spacing 
$\Delta x = 2\pi L/N_x $,  $\Delta y = 2\pi H/N_y $ and
$\Delta z = 2\pi H/N_z $ are equal where $N_x,N_y,N_z$ 
is the number of grid points in each direction, with $N_x=8N_y=8N_z$.

The simulations were started from the $\bu=0$ initial conditions
and continued until a steady state is reached for which a clear 
mean energy profile can be calculated. The only exception to this rule 
is the highest resolution run $N_x=8192$ for which the results of
the $N_x=4096$ run were extrapolated to a larger grid and used as
initial conditions. This run was performed for eight turn-over times
that was enough to converge sign-definite quantities (like energy)
but not sign-indefinite quantities (like fluxes).
A list with the properties of all runs performed are given in table \ref{tab:Re}.


\begin{table}
\begin{center}
  \begin{tabular}{cccc}
       \hspace{1cm} $ N_x\times N_y \times N_z$ \hspace{1cm} & 
       \hspace{1cm} $Re_\epsilon$                      \hspace{1cm} & 
       \hspace{1cm} $Re_U$               \hspace{1cm} & 
       \hspace{1cm} $Re_\lambda$                \hspace{1cm} \\[3pt]
       \hline
       $64 \times 64 \times 512 $     &    2.0   &    1.7     &   2.2    \\
       $64 \times 64 \times 512 $     &    4.0   &    4.5     &   5.8    \\
       $64 \times 64 \times 512 $     &    10.0  &    15.1    &   16.8   \\
       $64 \times 64 \times 512 $     &    20.0  &    34.7    &   31.3   \\
       $64 \times 64 \times 512 $     &    40.0  &    78.1    &   54.1   \\
       $128 \times 128 \times 1024 $  &    110   &    217     &   92.1   \\ 
       $256 \times 256 \times 2048 $  &    230   &    502     &  161     \\
       $512 \times 512 \times 4096 $  &    500   &    1165    &  270     \\
       $1024\times 1024 \times 8192 $ &   1250   &    2990    &  447     \\
  \end{tabular}
  \caption{ Resolution and values of the Reynolds numbers $Re_U,Re_\epsilon,Re_\lambda$ achieved in the numerical simulations. 
  }
  \label{tab:Re}
  \end{center}
\end{table}


\section{ Results}\label{sec:dns} 

\begin{figure}                                                                                   
  \centerline{\includegraphics[width=0.9\textwidth]{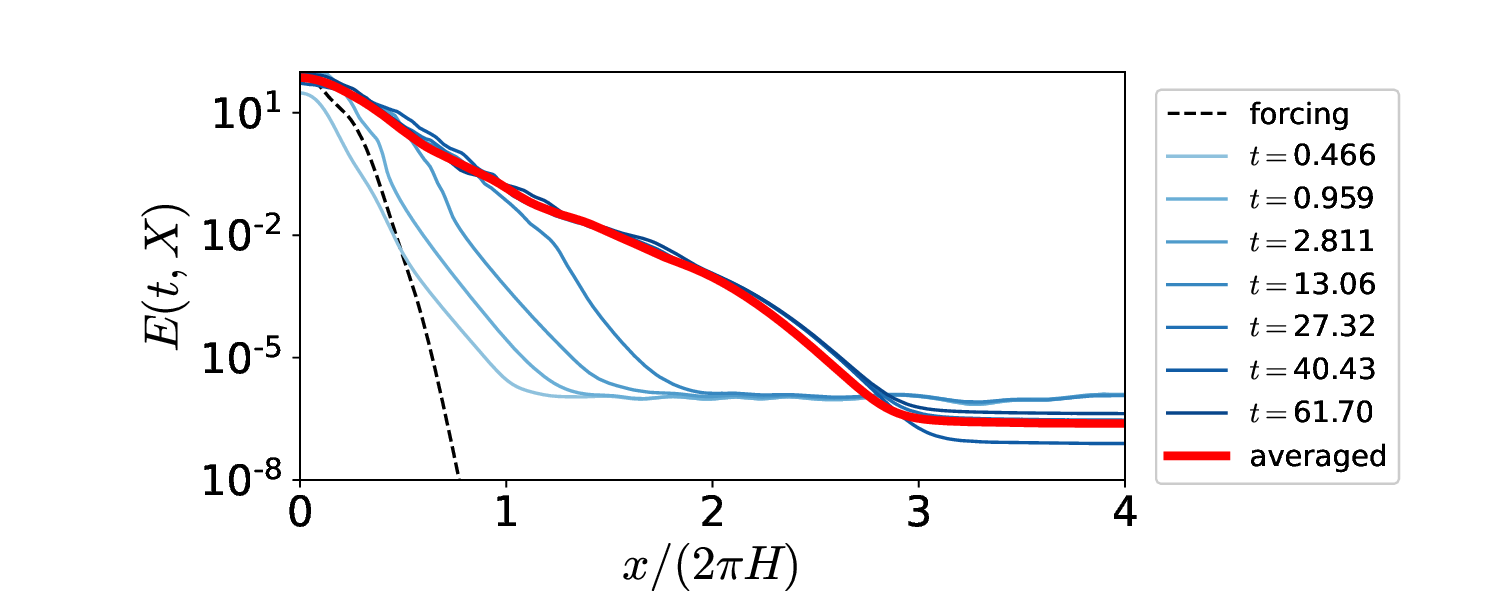} }
  \centerline{\includegraphics[width=0.9\textwidth]{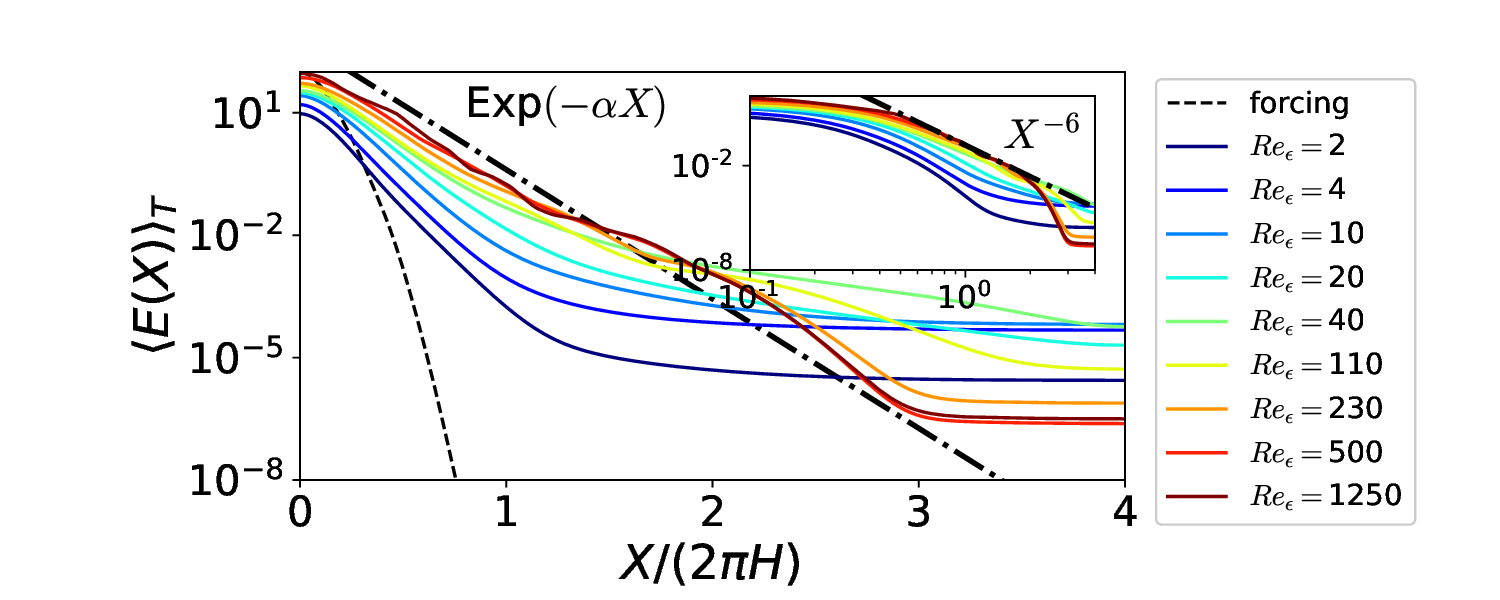} }
  \caption{
  Top panel: The energy density $E(t,X)$ for different times.  
  Bottom panel: The time averaged energy density $\langle E(X) \rangle_{_T}$ at steady state for different values of $Re_\epsilon$ in the entire domain.
  the dashed line indicates the forcing amplitude as a function of $X$. The inset shows the same data in log-log scale.   
  The same color index is used to mark $Re_\epsilon$ in all subsequent figures.}
\label{fig:energy}
\end{figure}

We begin with the top panel of figure \ref{fig:energy} that  shows the energy density $E(t,X)$ for $Re_\epsilon=500$ for different times. 
The black dashed line shows the forcing profile that is limited to $|X|/(2\pi H)\lesssim 1/2$.
Energy spreads away from the forcing region but at late times it fluctuates around a mean profile
shown by the red line. Thus already at this stage it can be testified that energy does not spread in the entire box and it remains close to the forcing region. 
This mean profile is shown in the bottom panel of the same figure for different values of $Re_\epsilon$.
The different colors indicate the different values of the Reynolds number achieved as marked in the
legend. The same colors are used for all subsequent figures.  The peak of the local energy density lies close to the forcing region 
and decays fast away from it. The energy far away from the forcing at $|X|/(2\pi H) \simeq 4$ remains very small such that $E(8\pi H)/E(0)\lesssim 10^{-6}.$
In the remaining of this section we will  try unravel the processes of this localisation and the implications for the system behavior.

\begin{figure}
  \centerline{
   \includegraphics[width=0.45\textwidth]{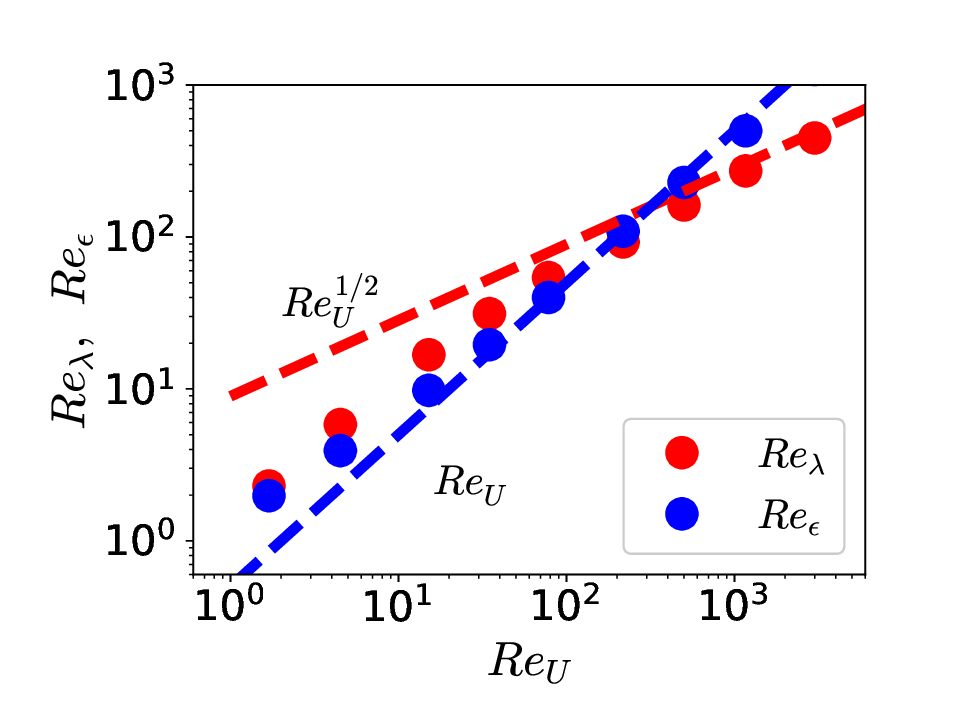}
   \includegraphics[width=0.45\textwidth]{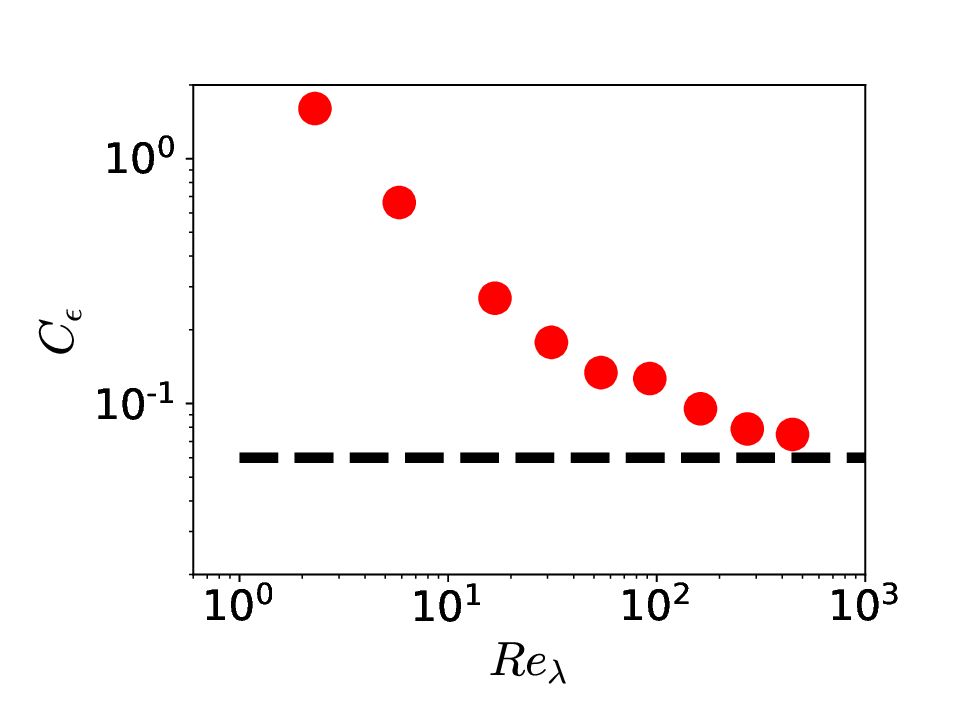} }
  \caption{ 
   Left: Relation between the different Reynolds numbers $Re_U,Re_\epsilon,Re_\lambda$.
   Right: The normalized dissipation rate $C_f$ as a function of $Re_\lambda$. }
\label{fig:C_f}
\end{figure}

Before continuing with spatial properties of our flow we perform 
some standard benchmark analysis often used in homogeneous turbulence. 
Figure \ref{fig:C_f}  shows the scaling of global measures as a function of the Reynolds number.
The left panel shows the relation between the different Reynolds numbers where the scaling $Re_U\propto Re_\epsilon \propto Re_\lambda^2$
that holds for large $Re$ is verified.
In right panel of figure \ref{fig:C_f} we show 
the non-dimensional dissipation rate (or drag coefficient) $C_\epsilon$ defined here as: 
\beq 
C_\epsilon = \frac{\epsilon H}{U^3}, 
\eeq 
that expresses the rate energy is dissipated non-dimensionalized by the amplitude of the fluctuations.
It is a corner-stone conjecture of homogeneous and isotropic turbulence theory that $C_\epsilon$ obtains a 
finite and $Re$-independent value at large $Re$. 
The present data  indicate that at large $Re_\lambda$, $C_\epsilon$ appears to converge to a $Re_\lambda$-independent value but quite slowly. 
Only the largest values of $Re_\lambda \gtrsim 270$ indicate the possibility that such a plateau is reached with a value of $C_\epsilon\simeq 0.06$ that is rather small.
In homogeneous and isotropic simulations such a plateau is reached after $Re_\lambda\sim 100$ and at a value much larger $C_\epsilon\simeq 0.5$ \citep{kaneda2003energy}.  
This reflects that localized turbulence is affected by the the additional freedom to expand in a larger region suppressing possibly its efficiency to cascade energy to the smaller scales.

\begin{figure}
  \centerline{\includegraphics[width=0.45\textwidth]{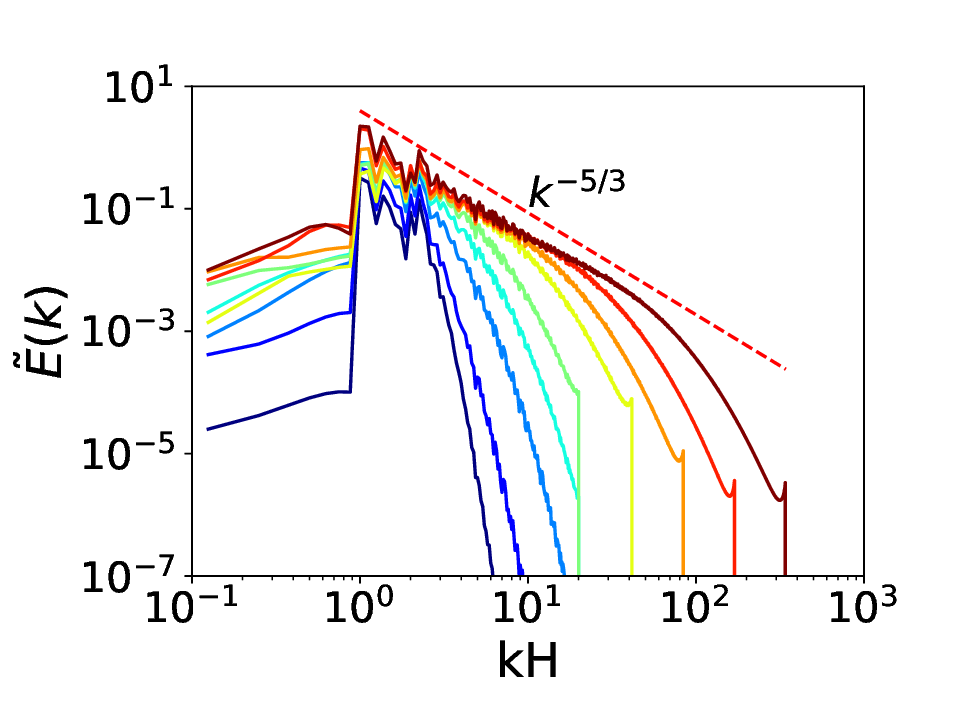}
              \includegraphics[width=0.45\textwidth]{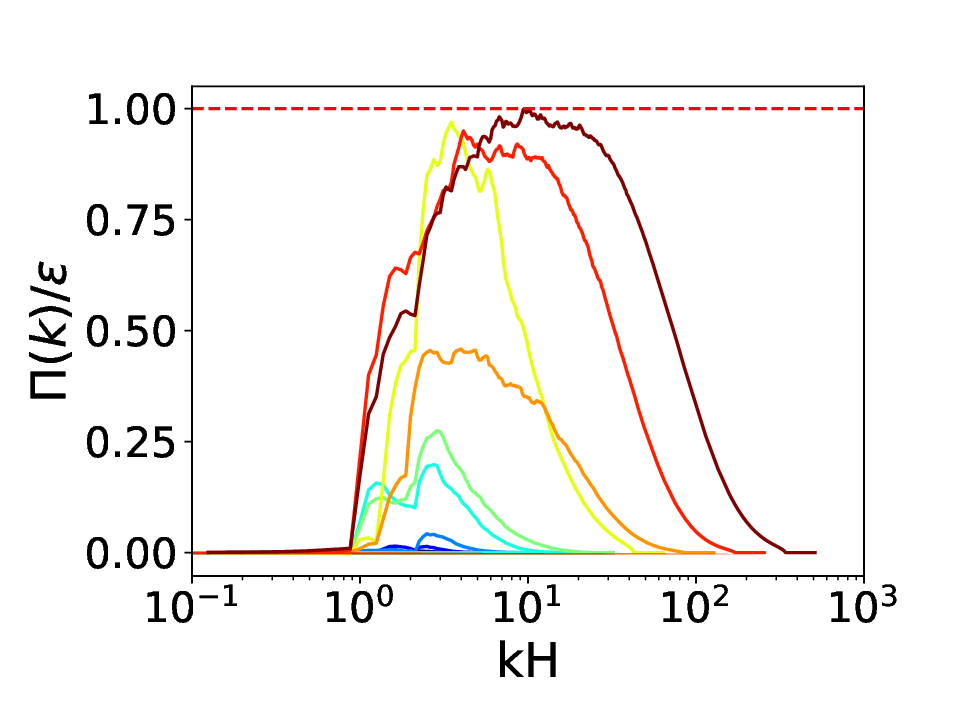} }
  \caption{ Left: The energy spectra $\tilde{E}(k)$ for the different $Re_\lambda$ examined.
            Right: The energy fluxes $\Pi(k)$ for the same runs. }
\label{fig:Espec}
\end{figure}

Figure \ref{fig:Espec} examines spectral properties of the flow.
In the left panel  we plot the energy spectra for the different values of $Re$. 
The spectra show similar behavior with homogeneous turbulence flows.
As the Reynolds number is increased more scales are excited and a power-law spectrum starts to form with exponent close to the Kolmogorov prediction ${\tilde{E(}}k)\propto k^{-5/3}$. 
%
In the right panel of figure \ref{fig:Espec} the energy fluxes in Fourier space
are plotted. The energy fluxes increase with $Re$ until for the largest $Re$s
attained a constant flux range has began to form. It is worth noting that 
this constant flux region is obtained at much larger $Re$ than what is observed 
in homogeneous turbulence simulations reflecting once again a delay in obtaining
a $Re$-independent scaling due to the effect of spreading.


Returning to the spatial properties of the flow and the energy density profile
we note that as the Reynolds number is increased the energy increases and also spreads at larger distances.
At very large values of $Re_\lambda$ the energy profile appears to converge to a $Re$ independent profile. 
This implies that at large $Re$ the energy profile is 
determined by the self-spreading of  eddies due to turbulent advection and not by viscous processes.
The fast drop of $E(X)$ can be either an exponential 
$E(x)\propto \exp(-\alpha x)$ of or fast power law $E(x)\propto |X|^{-6}$ (see inset).
The present data can not exclude either option. We point out that since
the energy density drops very fast also the local Reynolds number (defined using a local rms velocity) is also 
decreasing. So it is hard to obtain a large $Re$ behavior in the outer region $ |X| \gg H$. 

The fact that the energy density reaches a $Re$-independent profile is not a trivial result.
It reflects a balance between the rate energy is transported to larger values of $|x|$ and
the rate energy cascades to the small scales. If the cascade process was weaker than the real-space 
transport then at the $Re\to\infty$ limit energy would reach the entire domain. On the contrary 
if the real-space transport was weaker no energy would be found outside the forcing region 
in the same limit. In other words turbulent diffusion and turbulent dissipation  must be of the same order.

\begin{figure}
  \centerline{\includegraphics[width=0.33\textwidth]{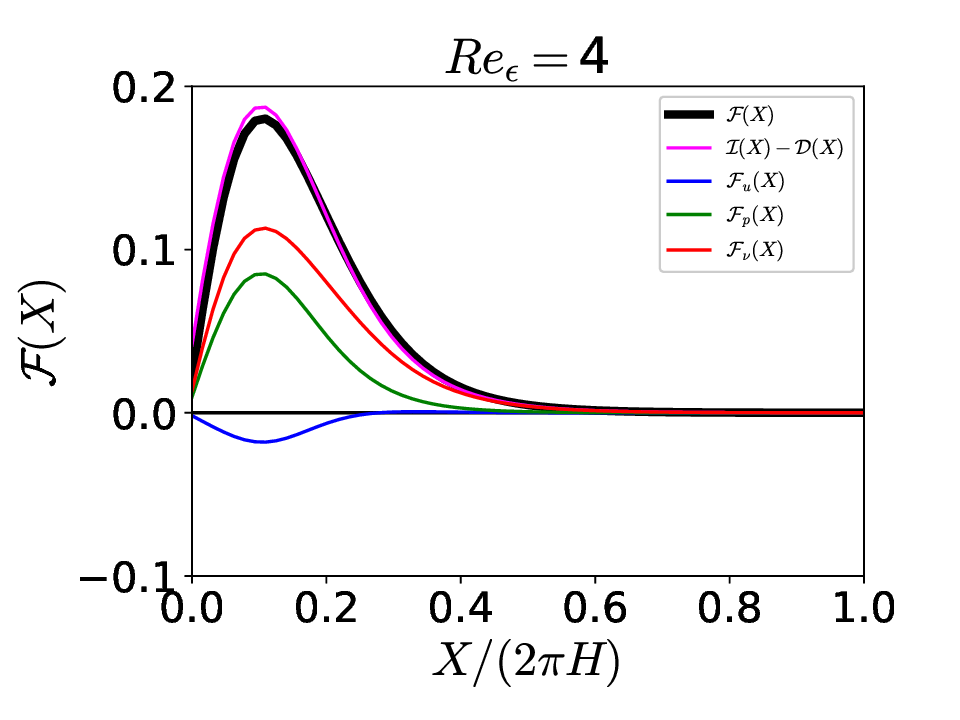},
              \includegraphics[width=0.33\textwidth]{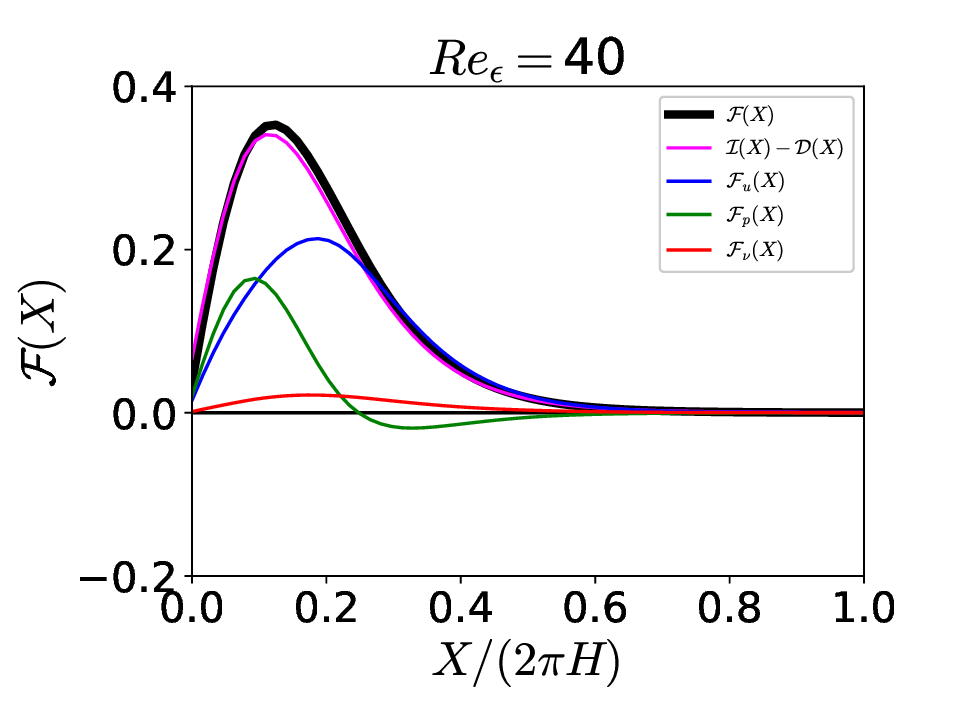},
              \includegraphics[width=0.33\textwidth]{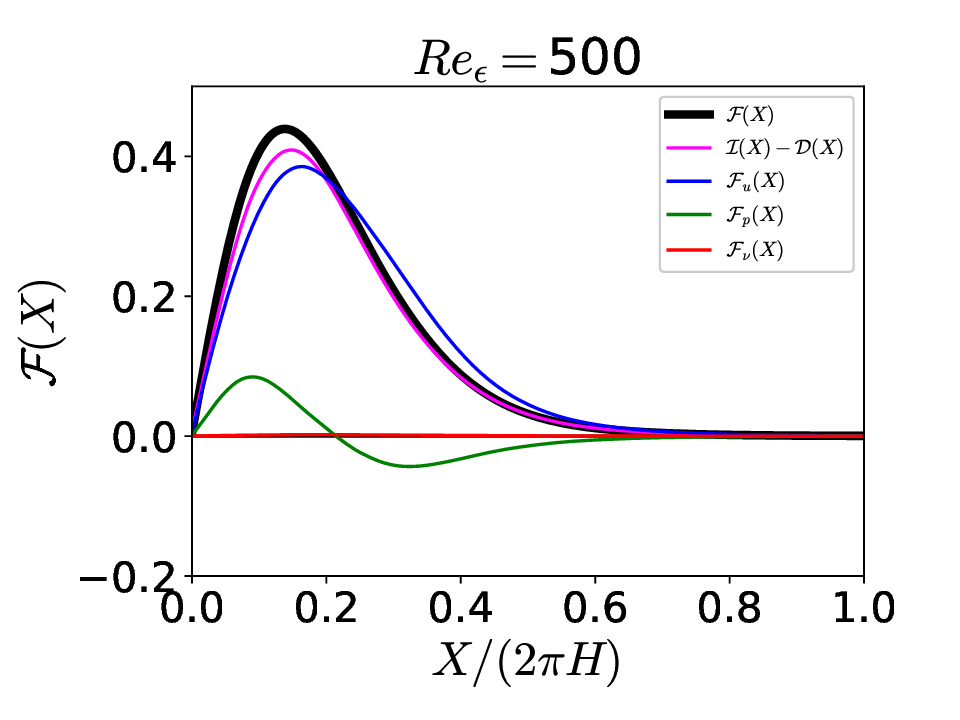}}
  \caption{ The different energy fluxes in real space as indicated in the legend for three different values of $Re_\epsilon=4,40,500$. }
\label{fig:fluxes}
\end{figure}

To quantify this assertion we look at the fluxes at real space.
In figure \ref{fig:fluxes} we plot $\cF_i$ for three different values of $Re$
varying from the laminar to the turbulence case. 
The black line shows the total flux, the blue line the flux due to velocity fluctuations, the green line the flux due to pressure and
the red line the flux due to viscosity. The magenta line shows the difference between  $\cI(x)$ and $\cD(x)$. 
A comparison between the back and magenta lines verifies 
the relation \ref{eq:fluxbal}. The small differences that are observed are due to insufficient 
time averaging that is more pronounced in the large resolution runs. 
A few observations need to follow.
For small $Re$ the energy flux is dominated by viscosity with pressure also playing a significant part.
The flux due to the velocity fluctuations have a negative sign.
%
As the Reynolds is increased the role of the velocity fluctuations becomes more dominant
transferring outwards energy. The transfer due to viscosity diminishes while the transfer due to pressure also takes negative values.  
At the largest $Re$ almost the entire flux is dominated by the velocity fluctuations with the pressure flux being weaker and 
positive in the forcing region and negative away from it.

\begin{figure}
  \centerline{\includegraphics[width=0.35\textwidth]{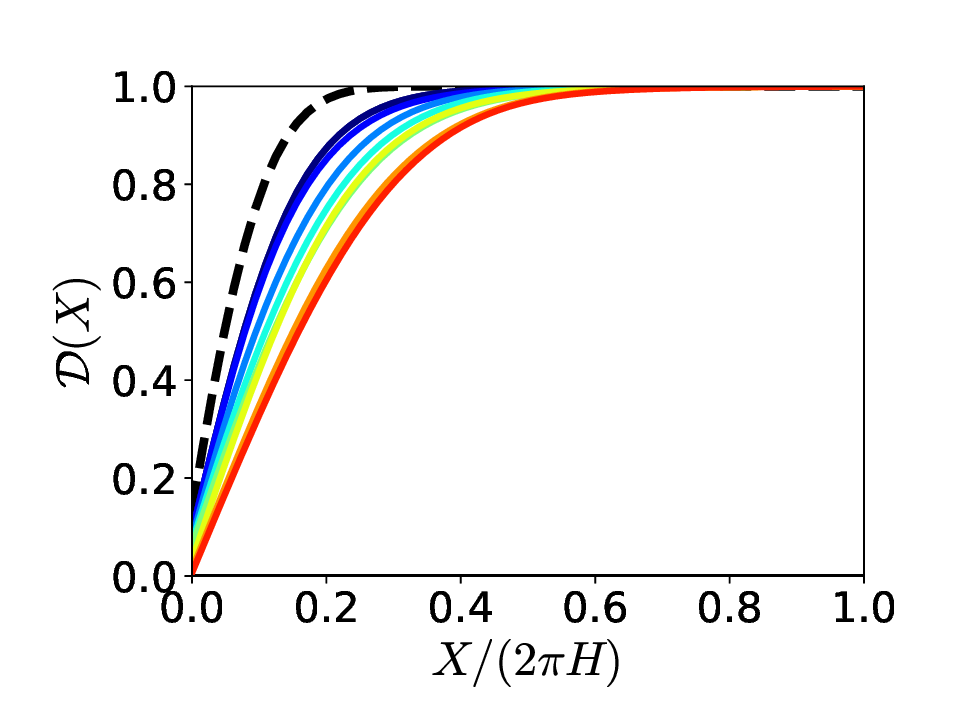},
              \includegraphics[width=0.35\textwidth]{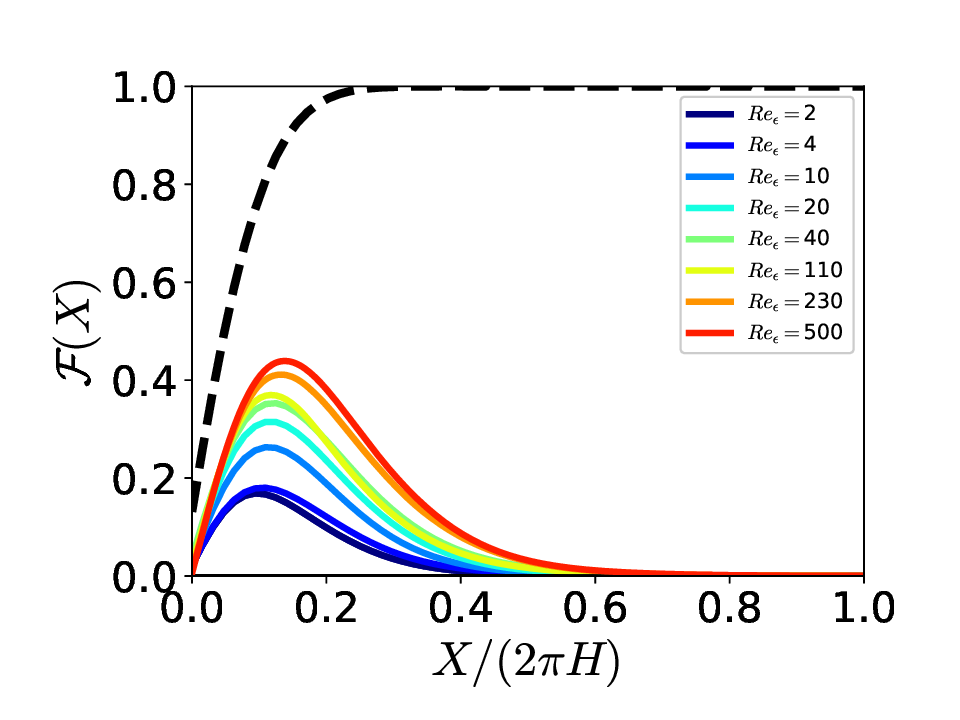},
              \includegraphics[width=0.35\textwidth]{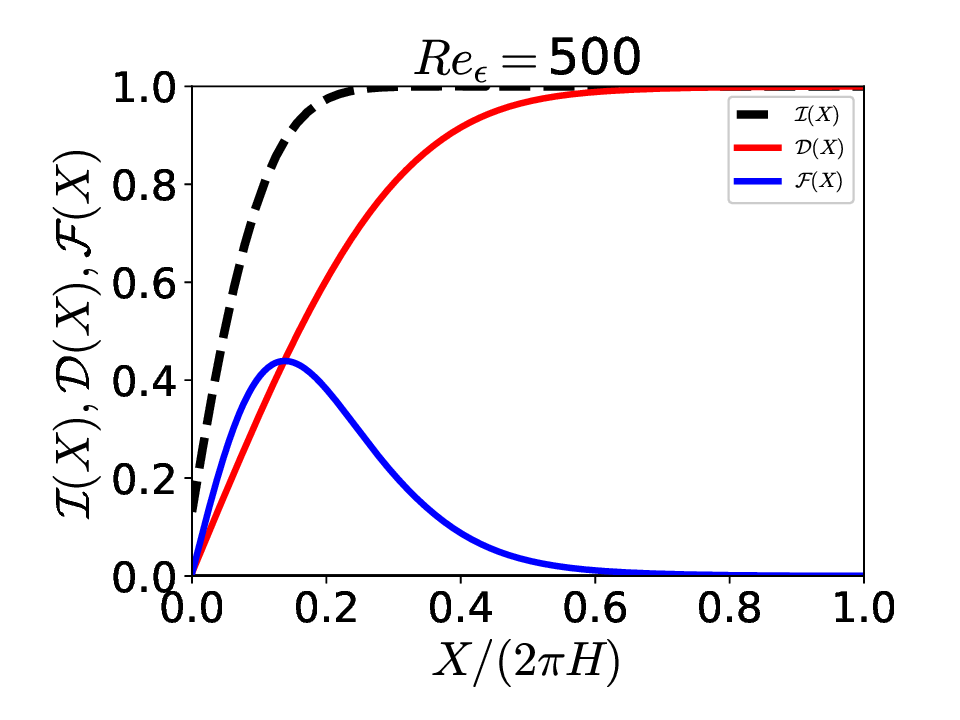}}
  \caption{ The dissipation rate $\cD(X)$ (left panel), the energy flux $\cF(X)$ (center panel) for different values of $Re$.
  The right panel compares the largest $Re_\epsilon=500$ for which the fluxes were measured. The dashed line indicates $\cI(X)$.  }
\label{fig:diss}
\end{figure}

Finally to compare the two dominant processes away from the forcing region the turbulent dissipation and the turbulent diffusion 
we plot in figure \ref{fig:diss} the dissipation rate $\cD(X)$
in the left panel and the total flux $\cF(X)$ on the center panels for all $Re$.
The right panel compares the two, for the largest value of $Re$ for which the fluxes were calculated $Re_\epsilon=500$.
The black dashed lines indicates $\cI(X)$ that is the same for all $Re$.
As the Reynolds is increased the dissipation is decreased while the flux is increased.
For the largest $Re$ at the peak of the flux around $X\simeq 0.15 (2\pi H)$ the two processes become of approximately equal
marking that the two processes turbulent dissipation and the turbulent diffusion are of the same order.

\section{ Conclusions}\label{sec:conclusions} 

The present work has demonstrated that locally forced turbulence will not spread throughout the domain
provided that there is no mean injection of linear or angular momentum. 
It will remain localised forming an energy density profile that is $Re$ independent in the 
large $Re$ limit. Away from the forcing region the two dominant effects are turbulent dissipation and 
turbulent diffusion that were found to be of the same order. 
The exact functional form of the energy profile could not be determined from the present simulations.
Theoretical investigations and modeling could give further insight to this problem.

To expand the understanding of the two involved processes, turbulent diffusion and turbulent dissipation,  
a simultaneous scale-space and real-space analysis would be required 
either by introducing local smoothing \citep{germano1992turbulence,aluie2009localness,eyink2009localness,alexakis2020local} or two point analysis and the K{\'a}rm{\'a}n–Howarth‐Monin–Hill (KHMH) equation \citep{hill2001equations,hill2002exact}. The latter has been used recently to study boundary driven flows \citep{apostolidis2022scalings,apostolidis2023turbulent} and wakes \cite{chen2022scalings,chen2021turbulence} where the role the inhomogeneous energy  injection from the mean flow was emphasised. In the present flow, there is no mean flow and the primary terms in balance are the inter-scale transfer rate and turbulent transport in physical space, both of which are forcing and viscosity independent. 
Thus a new state of turbulence is present where two inertial effects, the energy flux in scale space and in real space, compete.
%
The fact that these two dominant processes are viscosity-independent makes their modeling particular difficult 
as there is no simplifying limit where one term will dominate over the other. Careful parametrization  would be required so that the correct energy profile is captured. 

Finally we would like to add that  the present study was limited to a triple periodic channel flow limiting the spreading in only one direction. Its extension to larger domains where turbulence can spread in  two or in all three directions is far from trivial and would need to be examined separately. Here experimental investigations would become much more beneficial than
numerical simulations. 


\backsection[Funding]{This work was granted access to the HPC resources of GENCI-TGCC \& GENCI-CINES (Project No. A0130506421). This work has also been supported by the Agence nationale de la recherche (ANR DYSTURB project No. ANR-17-CE30-0004). }

\backsection[Declaration of Interests]{The authors report no conflict of interest.}

\bibliographystyle{jfm}
\bibliography{InHomogeneous}

\end{document}